\documentclass[twocolumn, prl, aps, amsfonts]{revtex4}
\usepackage{graphicx}
\usepackage{amsmath}

\newcommand{\ders}[2]{\frac{d^{2} #1}{d #2^{2}}}

\newcommand{\sech}{\operatorname{sech}}
\newcommand{\pf}[2]{\lp\frac{#1}{#2}\rp}
\renewcommand{\d}{\partial}
\newcommand{\unit}[1]{\text{ #1}}
\renewcommand{\v}[1]{\mathbf{#1}}

\newcommand{\lp}{\left(}
\newcommand{\rp}{\right)}
\newcommand{\lb}{\bigg[}
\newcommand{\rb}{\bigg]}
\newcommand{\mean}[1]{\langle #1 \rangle}
\renewcommand{\k}{\v{k}}
\newcommand{\x}{\vgreek{\varrho}}
\renewcommand{\r}{\v{r}}

\newcommand{\vgreek}[1]{\boldsymbol{#1}}

\begin{document}

\title{Anisotropic Instabilities in Trapped Spinor Bose-Einstein Condensates}
\author{M.~Baraban}
\author{H.~F.~Song}
\author{S.~M.~Girvin}
\author{L.~I.~Glazman}
\affiliation{Department of Physics, Yale University, New Haven, CT 06520}

\begin{abstract}
  We theoretically investigate the effect of an anisotropic trap on
  the instability of the polar $(m_F=0)$ phase of a spin-1 Bose-Einstein
  condensate. By considering rigorously the spatial quantization, we
  show that the growth of the nascent ferromagnetic phase at
  short times becomes anisotropic with stronger oscillations in the magnetization
  correlation function along the unconfined direction.
\end{abstract}

\maketitle

\setcounter{secnumdepth}{2} 

\section{Introduction}
Bose-Einstein condensates (BECs) with active spin degrees of freedom
have proven to be a fertile ground for studying cooperative quantum
many-body phenomena in condensed matter systems. Following the initial
realization of BECs with spin-1 alkali atoms in an optical trap
\cite{stamper-kurn:1} and description of the dynamics of the
system \cite{ho,ohmi}, theoretical and experimental investigations of
spinor BECs have identified such diverse behavior as quantum phase
transitions \cite{stamper-kurn:2,lamacraft}, domain and topological
defect formation \cite{stamper-kurn:2,saito:topological,mias,saito:kz},
coherent spin-mixing and amplification \cite{spin-mixing, number, amplifier}, and, most recently,
dipolar effects \cite{yi:dipolar, saito:dipolar, stamper-kurn:3, demler}.

In the experiment described in \cite{stamper-kurn:2}, a condensate of
${}^{87}$Rb atoms restricted to the ferromagnetic $F=1$ hyperfine
manifold was prepared in the $m_F=0$ state via the quadratic Zeeman
effect. When the magnetic field was rapidly quenched, quantum
fluctuations triggered the growth of unstable modes, leading to the
formation of domains of magnetization with random orientations in the
plane perpendicular to the spin-quantization axis. The instability is
characterized by the inverse time constant $\Gamma$ and corresponding
length scale $\xi=\sqrt{\hbar/2m\Gamma}$, which are set by the strength
of the spin-spin interaction [cf. Eq.~(\ref{eq:units})]. $\xi$ coincides
with the spin-healing length, with $\xi=2.4\unit{$\mu$m}$ for the
conditions of Ref.~\cite{stamper-kurn:2}.

The exponential growth of the amplitude of a mode is proportional to
$e^{\Omega\Gamma t}$ and is controlled by the dimensionless gain
parameter $\Omega$. In a homogeneous system, an unstable mode may be
characterized by a wavevector $\k$. Neglecting the dipolar interaction, one finds \cite{lamacraft} the gain parameter to be $\Omega=\sqrt{k^2(2-k^2)}$, where $k$ is measured in units of $\xi^{-1}$.
Although the nature of the instability at early times ($\Gamma t
\lesssim 4$) has been successfully analyzed in the case of an infinite
condensate \cite{saito:kz,mias,lamacraft}, the effect of the trap
geometry on the instability and spatial distribution of the associated
magnetization has not been adequately addressed. This is a relevant
question, since the experiment observed significant anisotropy in the
magnetization correlation function which was attributed to the spatial
anisotropy of the condensate. It was argued that as a result of the
elongated shape of the condensate a small set of discrete modes in the
trapped direction, rather than a continuum of states, led to
preferential modulation of the magnetization in the confined direction.

In this paper, we account rigorously for the spatial quantization effects in the development of the instability of the polar phase. Concentrating on the case of isotropic interactions within the $F=1$ manifold, we find that the correlation function [Eq.~(\ref{eq:Gfirst})] displays behavior precisely opposite to the one observed in experiment, i.e., displays a more prominent magnetization modulation in the unconfined direction. The inconsistency may be due to distortions introduced by optical aberrations in the imaging device, and new experiments are underway \cite{stamper-kurn:pc}.

\begin{figure}[htp]
\includegraphics[width=3.25in]{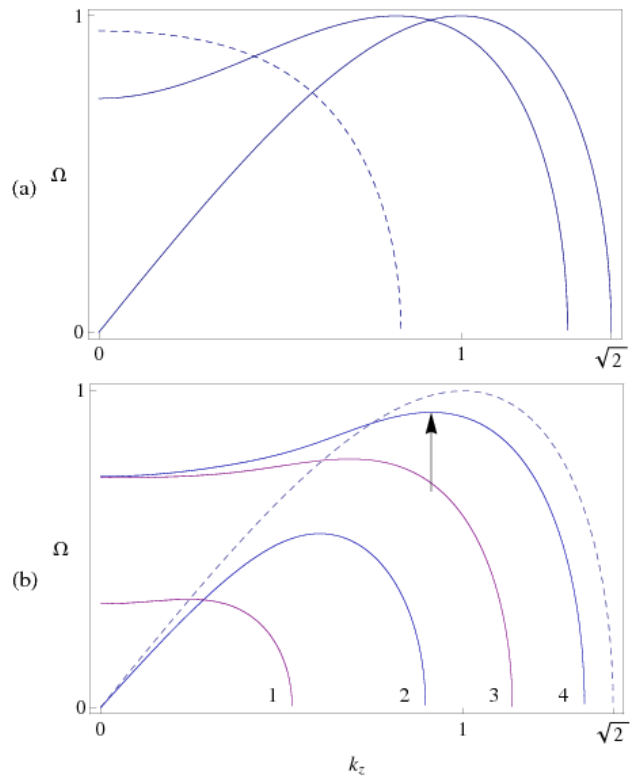}
\caption
{\label{fig:gain} (color online). Comparison of the gain parameter $\Omega(k_z)$ for a two-dimensional spinor BEC in the $x-z$ plane, for different models of confinement in the $x$-direction. (a) Periodic boundary conditions (the simplest model) and (b) realistic confinement in a harmonic trap potential; $L_x/\xi = 11$ in both cases. In (a) the solid and dashed curves correspond to $k_x <1$ and $k_x>1$, respectively, and the two modes with $\Omega\neq0$ at $k_z=0$ are each doubly-degenerate. Note that all three modes with $k_x<1$ have the same maximum value of gain $\Omega$. In (b), curves of the same parity (even for 1 and 3, odd for 2 and 4) show anti-crossing behavior due to mode-mixing, especially evident in the latter pair. The position of the arrow indicates the value of $k_z$ corresponding to the maximum gain; the instability with this gain may become dominant before depletion becomes significant, see Sec.~IIID. The dashed curve in (b) is the $k_x = 0$ curve from (a), for comparison.}
\end{figure}

To be specific, we consider a condensate whose thickness in the
$y$-direction is comparable to or smaller than the spin-healing length
so that the spin dynamics is effectively constrained to the $x-z$
plane. Moreover, we assume that the condensate
is infinite in the $z$-direction but short in the $x$-direction due to
a confining potential. Note that the effect of a trap on
single-component condensates has already been analyzed extensively
\cite{pethick}. However, the analysis of a condensate with a spinor
order parameter is complicated by the fact that, in the case of isotropic interactions within the $F=1$ manifold, the $SU(2)$ symmetry in
spin space must be respected. In particular, the simplest approach in
which we keep the functional form of $\Omega(\k)$ for an infinite system but restrict the allowed wavevectors to certain values by applying either hard-wall boundary conditions or periodic boundary conditions (PBCs) yields qualitatively incorrect results. In the case of hard-wall boundary conditions, $k_x$ takes on discrete values which exclude $k_x=0$, which results in $\Omega\neq0$ for all allowed values of $\k$ and contradicts the existence of a zero mode in the $SU(2)$-symmetric case \cite{lamacraft}. Meanwhile, although the presence of the
zero mode is preserved by choosing PBCs in the $x$-direction, this results in the wrong
systematics of unstable modes (only an odd number of such modes are
possible, and the maximal gain is identical for all modes with
$k_x\leq1$). The errors introduced by either of these two
simplifications are insignificant for wide traps ($L_x\gg\xi$), but
grow in importance with decreasing $L_x$. For the typical trap widths
of the order of a few $2\pi\xi$, these errors are substantial enough to
affect the qualitative features of the instability development. As an
illustration, we plot in Fig.~\ref{fig:gain} the gain parameter
$\Omega(k_z)$ for a system with only a few modes spatially quantized
in the $x$-direction by (a) PBCs and (b) a rigorous consideration of a harmonic trap. In the former case there are 2 (3, accounting for $\pm k_x$ degeneracy) modes with equal
maximum gain and another mode with a nearly-degenerate maximum. The exact degeneracy of the different maxima for $k_x\leq1$ arises because the differences in $k_x$ are compensated by different values of $k_z$ satisfying $k^2=k_x^2+k_z^2=1$. In contrast, the latter is characterized by substantial differences between the maxima of $\Omega$ for different modes. In time, this leads to an
increasingly pronounced anisotropy of the magnetization correlation
function because a \emph{single} mode with the largest gain dominates the growth of the nascent ferromagnetic phase. Since this mode has a well-defined wavelength in the $z$-direction, increasingly strong oscillations between regions of positive and negative correlations are expected in the $z$-direction. Note that when $L_x\lesssim 2\pi\xi$ only a single mode corresponding to $k_x=0$ is unstable and applying PBCs makes qualitatively correct predictions for the behavior of the magnetization correlation function. Indeed, this justifies our later treatment of the $y$-direction in a single-mode approximation, since $L_y< 2\pi\xi$.

In the following section we solve the equations of motion for the
amplitude of the confined $m_F=\pm1$ condensate during the exponential
growth stage. On the basis of these solutions, in Sec.~III we explore
the anisotropy of the resulting magnetization correlation function. In
doing so we emphasize the importance of the zero mode guaranteed by
$SU(2)$ symmetry, both to the spectrum of instabilities and to the
stability of a spinor BEC in the absence of any spin-spin interaction.

\section{Short-time dynamics}
\subsection{Equations of motion}
In the absence of magnetic fields, an interacting spin-1 Bose gas is
described by the Hamiltonian \cite{ho}
\begin{align}
        \hat{H} &= \int d\r\ :\hat{{\cal H}}:\label{eq:Ham}, \\
        \hat{{\cal H}} &= \hat{\Psi}^\dag\lb -\frac{\hbar^2}{2m}\nabla^2 + V - \mu\rb\hat{\Psi}
        + \frac{c_0}{2}\hat{n}^2
        + \frac{c_2}{2}\v{\hat{F}}^2\label{eq:fullH},
\end{align}
where $\hat{\Psi}$ is a spinor whose components
$\hat{\Psi}_{m_F}^\dagger$ create particles in magnetic sublevel
$m_F=+1,0,\text{ or }-1$, $\hat{n}=\hat{\Psi}^\dag\hat{\Psi}$,
$\v{\hat{F}}=\hat{\Psi}^\dag\v{F}\hat{\Psi}$ with $\v{F}$ being the
spin-1 matrices quantized along the $z$-axis, $V(\r)$ is the trap
potential, and $\mu$ is the chemical potential. The interaction
constants are related to the $s$-wave scattering lengths $a_F$ by
$c_0=4\pi\hbar^2(a_0+2a_2)/3m,\ c_2=4\pi\hbar^2(a_2-a_0)/3m$, and we
assume that the spin-spin interaction is ferromagnetic ($c_2<0$) as in
${}^{87}$Rb. To find the evolution of the observables of the initially
polar ($m_F=0$) state, we may impose two simplifications based on a
zero-depletion approximation whose domain of validity we consider
later in the paper: 1) keep the $m_F=\pm1$ components to lowest order,
i.e., to quadratic order in the Hamiltonian, and 2) treat
$\hat{\Psi}_0$ as a classical real field $\psi_0(\r)$ satisfying the
ordinary Gross-Pitaevskii equation for a scalar condensate:
\begin{equation}
        \lb -\frac{\hbar^2}{2m}\nabla^2 + V
        + c_0\psi_0^2-\mu \rb \psi_0 = 0\label{eq:scalargp}.
\end{equation}
With these assumptions the Heisenberg equations of motion yield a closed system for
$\hat{\Psi}_{+1},\hat{\Psi}^\dag_{-1}$:
\begin{align}
        i\hbar\d_t\hat{\Psi}_{+1} &= (H_0 + c_2n)\hat{\Psi}_{+1} + c_2n\hat{\Psi}_{-1}^\dag\label{eq:hei1},\\
        -i\hbar\d_t\hat{\Psi}^\dag_{-1} &= (H_0 + c_2n)\hat{\Psi}^\dag_{-1} + c_2n\hat{\Psi}_{+1}\label{eq:hei2}.
\end{align}Here $n=\psi_0^2$ and, using Eq.~(\ref{eq:scalargp}),
\begin{align}
        H_0 &= -\frac{\hbar^2}{2m}\nabla^2 + V + c_0n - \mu \\
        &= \frac{\hbar^2}{2m}\lb -\nabla^2 + \pf{\nabla^2\psi_0}{\psi_0}\rb\label{eq:H0n}.
\end{align}
Note that the latter form for $H_0$ guarantees that $\psi_0$ is an
eigenstate of $H_0$ with zero eigenvalue. Furthermore, since $\psi_0$ satisfying Eq.~(\ref{eq:scalargp}) is nodeless, the ground state energy and all other eigenvalues of $H_0$ are non-negative. As we
will see later, this property is essential to respecting
spin-conservation in this system in the absence of magnetic fields.

Another important consequence is that a spinor BEC in the absence of
spin-spin interactions is stable for any homogeneous spin state, an expected yet
non-trivial fact. Although this is already clear from Eqs.~(\ref{eq:hei1}),(\ref{eq:hei2}) with $c_2=0$, it is worth seeing this explicitly from the multi-component Gross-Pitaevskii equation corresponding to Eq.~(\ref{eq:fullH}) with $c_2=0$:
\begin{equation}
        i\hbar\d_t\Psi = \lb -\frac{\hbar^2}{2m}\nabla^2 + V + c_0|\Psi|^2 - \mu\rb\Psi.
\end{equation}
Because of the (spin-) rotational symmetry, the solution will be of
the form $\Psi_0(\r)=\psi_0(\r)\zeta$, where $\zeta$ is a
position-independent spinor. This allows us to decompose small fluctuations about the mean-field as $\delta\Psi=\delta\Psi_\parallel+\delta\Psi_\perp$, where
$\delta\Psi_\parallel$ is along $\zeta$ and
$\Psi_0^\dag\delta\Psi_\perp=\delta\Psi^\dag_\perp\Psi_0=0$, so that
\begin{equation}
        i\hbar\d_t\delta\Psi_\perp = \lb -\frac{\hbar^2}{2m}\nabla^2 + V + c_0\psi_0^2 - \mu\rb\delta\Psi_\perp
        = H_0\delta\Psi_\perp.
\end{equation}
Here $H_0$ is the same Hamiltonian from Eq.~(\ref{eq:H0n}). The fact
that $H_0$ is bounded from below by zero therefore guarantees that in the absence of spin interaction, i.e., $c_2=0$, the ground state energy of the BEC for any position-independent spin state (characterized by spinor $\zeta$) is the same, and coincides with the one for a scalar BEC.

Returning to the linearized quantum equations of motion with $c_2<
0$, the density at the center of the condensate, $n_0$, defines a
characteristic inverse time constant for the instability and corresponding length
\begin{equation}
        \Gamma = \frac{|c_2|n_0}{\hbar}, \qquad \xi = \sqrt{\frac{\hbar}{2m\Gamma}}\label{eq:units}
\end{equation}
for the spin-spin interaction. It will be convenient to measure times
and distances in terms of $\Gamma^{-1}$ and $\xi$, respectively; we
will use the
notation
\begin{equation}
\tau = \Gamma t,\qquad \x = \frac{\r}{\xi},\qquad \hat{\Phi}_{\pm1} =
\xi^{3/2}\hat{\Psi}_{\pm1}
\end{equation}
for the dimensionless quantities. We then arrive at the system of equations
\begin{align}
        i\d_\tau\hat{\Phi}_{+1} = (h_0 - \rho)\hat{\Phi}_{+1} - \rho\hat{\Phi}_{-1}^\dag,\label{eq:p1}\\
        -i\d_\tau\hat{\Phi}^\dag_{-1} = (h_0 - \rho)\hat{\Phi}^\dag_{-1} - \rho\hat{\Phi}_{+1}\label{eq:m1}
\end{align}where $h_0=H_0/\hbar\Gamma$:
\begin{equation}
        h_0 = -\nabla^2 + \pf{\nabla^2\sqrt{\rho}}{\sqrt{\rho}},\qquad
	\rho=\frac{n}{n_0}=\frac{\psi_0({\r})^2}{n_0}.
        \label{eq:H0}
\end{equation}
The observable quantity of interest is the transverse magnetization
correlation function, which we define using $\hat{F}_\pm=\hat{F}_x\pm
i\hat{F}_y$ as
\begin{align}
        G_\perp(\r,\r',t) &=
        \mean{\hat{F}_+(\r,t)\hat{F}_-(\r',t)}\label{eq:Gfirst}\\
&= n_0^2\cdot \frac{1}{\xi^3n_0}g_\perp(\x,\x',\tau),
\end{align}
where the dimensionless correlation function is given by
\begin{align}
 &g_\perp(\x,\x',\tau)= 2\sqrt{\rho(\x)\rho(\x')}  \label{eq:littleg}\\
&\qquad\times\langle\hat{\Phi}_{+1}^\dag(\x,\tau)\hat{\Phi}_{+1}(\x',\tau)
+\hat{\Phi}_{+1}^\dag(\x,\tau)\hat{\Phi}^\dag_{-1}(\x',\tau)\notag\\
&\qquad\qquad+\hat{\Phi}_{-1}(\x,\tau)\hat{\Phi}_{+1}(\x',\tau)
+\hat{\Phi}_{-1}(\x,\tau)\hat{\Phi}^\dag_{-1}(\x',\tau)\rangle\notag
\end{align}
and expectation values are taken with respect to the initial
$m_F=\pm1$ vacuum state. In the following sections we derive the
complete solution for $g_\perp$ from which the homogeneous solution
for $V=0$ follows trivially, then specialize to the case where the
condensate has a finite extent in one direction.

\subsection{General solution for the magnetization correlation function}
The solution of the coupled linear system in Eqs.~(\ref{eq:p1}),(\ref{eq:m1}) can be written as
\begin{widetext}
\begin{align}
        \hat{\Phi}_{+1}(\x,\tau) &= \int d\x' \ \bigg[ U(\x,\x' ,\tau)\hat{\Phi}_{+1}(\x' ,0)
        + V^\ast(\x,\x' ,\tau)\hat{\Phi}_{-1}^\dag(\x' ,0)\bigg]\label{eq:p1s},\\
        \hat{\Phi}^\dag_{-1}(\x,\tau) &= \int d\x' \ \bigg[ V(\x,\x' ,\tau)\hat{\Phi}_{+1}(\x' ,0)
        + U^\ast(\x,\x' ,\tau)\hat{\Phi}_{-1}^\dag(\x' ,0)\bigg]\label{eq:m1s},
\end{align}
\end{widetext}with the initial conditions $U(\x,\x',0) = \delta(\x-\x')$, $V(\x,\x',0) = 0$. We may expand $U$ and $V$ in an orthonormal basis $\varphi_\alpha$ as
\begin{align}
        U(\x,\x',\tau) &= \sum_{\alpha,\beta}\varphi_\alpha(\x)u_{\alpha\beta}(\tau)\varphi^\ast_\beta(\x')\label{eq:U},\\
        V(\x,\x',\tau) &= \sum_{\alpha,\beta}\varphi_\alpha(\x)v_{\alpha\beta}(\tau)\varphi^\ast_\beta(\x')\label{eq:V}
\end{align}where $u_{\alpha\beta}(0) = \delta_{\alpha\beta}, \ v_{\alpha\beta}(0)=0$, while
\begin{align}
        \hat{\Phi}_{+1}(\x',0) &= \sum_\gamma \varphi_\gamma(\x')\hat{a}_{+1,\gamma},\\
        \hat{\Phi}^\dag_{-1}(\x',0) &= \sum_\gamma \varphi_\gamma^\ast(\x')\hat{a}^\dag_{-1,\gamma}.
\end{align}Then Eqs.~(\ref{eq:p1s}),(\ref{eq:m1s}) become
\begin{align}
\hat{\Phi}_{+1}(\x,\tau) &= \sum_{\alpha,\beta}\bigg[ \varphi_\alpha(\x)u_{\alpha\beta}(\tau)\hat{a}_{+1,\beta}
        +  \varphi^\ast_\alpha(\x) v^\ast_{\alpha\beta}(\tau)\hat{a}^\dag_{-1,\beta}  \bigg]\label{eq:Phip1},\\
        \hat{\Phi}^\dag_{-1}(\x,\tau) &= \sum_{\alpha,\beta}\bigg[ \varphi_\alpha(\x)v_{\alpha\beta}(\tau)\hat{a}_{+1,\beta}
        +  \varphi^\ast_\alpha(\x) u^\ast_{\alpha\beta}(\tau)\hat{a}^\dag_{-1,\beta}  \bigg]\label{eq:Phim1}.
\end{align}
Let $\varphi_\alpha$ be the eigenstates of $h_0$ in Eq.~(\ref{eq:H0})
with eigenvalues $\epsilon_\alpha$. Substituting this expansion into
Eqs.~(\ref{eq:p1}),(\ref{eq:m1}) and introducing the linear combinations
\begin{equation}
        w_{\alpha\gamma} = u_{\alpha\gamma}+v_{\alpha\gamma},
        \qquad z_{\alpha\gamma} = u_{\alpha\gamma}-v_{\alpha\gamma}\label{eq:lincomb}
\end{equation}then gives
\begin{align}
        i\dot{z}_{\alpha\gamma} &= \epsilon_\alpha w_{\alpha\gamma} - 2\sum_\beta\rho_{\alpha\beta}w_{\beta\gamma}\label{eq:z},\\
        i\dot{w}_{\alpha\gamma} &= \epsilon_\alpha z_{\alpha\gamma}\label{eq:w}.
\end{align}Here $\rho_{\alpha\beta} = \int d\x\ \varphi_\alpha^\ast(\x)\rho(\x)\varphi_\beta(\x)$ are the matrix elements of $\rho(\x)$. The corresponding initial conditions are
\begin{equation}
        w_{\alpha\gamma}(0) = z_{\alpha\gamma}(0) = \delta_{\alpha\gamma}\label{eq:init}.
\end{equation}
This is a particularly convenient formulation of the problem, since
the dimensionless magnetization correlation function in Eq.~(\ref{eq:littleg}) is easily shown, with the help of Eqs.~(\ref{eq:Phip1}),(\ref{eq:Phim1}) and
  the relations $\mean{a_{m\alpha}a^\dagger_{m'\beta}}=\delta_{mm'}\delta_{\alpha\beta}$, to be
\begin{align}
g_\perp(\x,\x',\tau)&=2\sqrt{\rho(\x)\rho(\x')}\nonumber\\
&\times\sum_{\alpha,\beta,\gamma}\varphi_\alpha(\x)\varphi_\beta^\ast(\x')
w_{\alpha\gamma}(\tau)w^\ast_{\beta\gamma}(\tau)\label{eq:gperp}.
\end{align}
Hence we are primarily interested in computing $w_{\alpha\gamma}$.

Now, if we na\"{i}vely try to solve Eqs.~(\ref{eq:z}),(\ref{eq:w}) by
taking the time derivative of Eq.~(\ref{eq:w}) and substituting Eq.
(\ref{eq:z}), we get
\begin{equation}
        \ddot{w}_{\alpha\gamma} = -\sum_\beta M_{\alpha\beta}w_{\beta\gamma}
\end{equation}
where $M_{\alpha\beta} = \epsilon_\alpha^2 \delta_{\alpha\beta} -
2\epsilon_\alpha\rho_{\alpha\beta}$ is non-hermitian. Thus it is
desirable to apply a transformation that renders the system hermitian.
But we may note now that one mode of the system will be neither
oscillating (positive eigenvalues of $M$) nor growing (negative
eigenvalues of $M$). This follows from the earlier-noted existence of
a mode we label as $\alpha=0$ such that $\epsilon_0=0$, which implies
$M_{0\beta}=0$. Since the first row contains only zeros, the
determinant of $M$ itself must be zero, which shows that zero is an
eigenvalue of $M$. This is an important consequence of the $SU(2)$
symmetry of the system \cite{lamacraft}; specifically, the
infinite-wavelength excitation creating an $m_F=\pm1$ pair is
equivalent to a global rotation of the system, and hence does not cost
any energy.

By performing the similarity transformation $\tilde{w}_{\alpha\gamma}=\epsilon_\alpha^{-1/2}w_{\alpha\gamma},\ \tilde{z}_{\alpha\gamma}=\epsilon^{1/2}_\alpha z_{\alpha\gamma}$ for $\alpha\neq0$, we find that the solution of Eqs.~(\ref{eq:z}),(\ref{eq:w}) is governed by the hermitian matrix 
\begin{equation}
        M_{\alpha\beta} = \epsilon_\alpha^2 \delta_{\alpha\beta} - 2\epsilon_\alpha^{1/2}\rho_{\alpha\beta}\epsilon_\beta^{1/2}, \qquad \alpha,\beta\neq0 \label{eq:M}.
\end{equation}
Let $S$ be the unitary matrix that diagonalizes $M$, $[S^\dag
MS]_{\lambda\lambda'} = E_\lambda\delta_{\lambda\lambda'}$. Some of
the eigenvalues $E_\lambda$ are negative, representing the unstable
modes of the system. Since the oscillatory modes for which $E_\lambda$
is positive are quickly washed out by the exponential growth of the
unstable modes, we may restrict all summations over $\lambda$ to
negative $E_\lambda$. For these we define
$\Omega_\lambda=\sqrt{-E_\lambda}$, which we call the ``gain'' in
reference to the connection with parametric amplifiers in quantum
optics \cite{walls}. Then
\begin{widetext}
\begin{align}
        w_{0\gamma}(\tau) &= \delta_{0\gamma}\label{eq:w1},\\
        w_{\alpha\neq0,0}(\tau) &= \sum_\lambda \sum_{\beta\neq0} S_{\alpha\lambda}S^\dag_{\lambda\beta}(\epsilon_\alpha\epsilon_\beta)^{1/2}\rho_{\beta0}\lp\frac{\sinh \Omega_\lambda \tau/2}{\Omega_\lambda/2}\rp^2\label{eq:w2},\\
        w_{\alpha\neq0,\gamma\neq0}(\tau) &= \sum_\lambda S_{\alpha\lambda}S^\dag_{\lambda\gamma}\lb
        \pf{\epsilon_\alpha}{\epsilon_\gamma}^{1/2} \cosh \Omega_\lambda \tau - i(\epsilon_\alpha\epsilon_\gamma)^{1/2}\frac{\sinh \Omega_\lambda \tau}{\Omega_\lambda}\rb\label{eq:w3}.
\end{align}
\end{widetext}

In summary, the transverse magnetization correlation function is
determined by Eq.~(\ref{eq:gperp}) through the eigenstates
$\varphi_\alpha$ of Eq.~(\ref{eq:H0}) together with Eqs.~(\ref{eq:w1})-(\ref{eq:w3}). 

Note that previous results for the homogeneous case follow trivially
from this formalism. If $V=0$ then $n=n_0$ is the solution to Eq.~(\ref{eq:scalargp}) and the eigenvalues and eigenstates of $h_0$ [Eq.~(\ref{eq:H0})] are $\epsilon_\k = k^2$ and $\varphi_\k(\x) \propto
e^{i\k\cdot\x}$, respectively. Since $\rho$ and $M$ are
diagonal, $S$ is simply the identity matrix and the gain function is
given by
\begin{equation}
        \Omega_\k = \sqrt{k^2(2-k^2)}\label{eq:omegak}.
\end{equation}Eqs.~(\ref{eq:w1})-(\ref{eq:w3}) and Eq.~(\ref{eq:gperp}) then give $g_\perp(\x,\x',\tau)=g_\perp(\x-\x',\tau)$ where
\begin{equation}
        g_\perp(\x,\tau) \propto \sum_{k^2\leq2} \left| \cosh \Omega_\k\tau - i\epsilon_\k\frac{\sinh \Omega_\k\tau}{\Omega_\k} \right|^2 e^{i\k\cdot\x}\label{eq:2dgperp},
\end{equation}
which is consistent with \cite{saito:kz}.

The formalism developed above is applicable to any trap geometry. In
the case of tight confinement in one of the directions (the $y$-direction, for concreteness, where $L_y\lesssim 2\pi\xi$), one may use a single-mode approximation for the $y$-dependence of eigenfunctions $\varphi_\alpha$,
as was done in previous works \cite{mias}. In the following sections we take into account the variation of $\varphi_\alpha$ with $x$ and $z$ only, so that we may constrain the problem to the solution of
differential equations in the $x-z$ plane.

\section{Anisotropic Confinement}

\subsection{Motivation}

To gain an idea of what differences we should expect in the anisotropic case, note that for sufficiently large times we can approximate Eq.~(\ref{eq:2dgperp}) as
\begin{equation}
        g_\perp(\x,\tau) \sim \sum_{k^2\leq2} \frac{e^{2\Omega_\k\tau+i\k\cdot\x}}{2-k^2}\label{eq:g2d}.
\end{equation}Carrying out the sum using steepest-descent integration around the maximum gain at $k=1$ gives (in strictly two dimensions)
\begin{equation}
        g_\perp(\x,\tau) \sim \frac{e^{2\tau}}{\sqrt{\tau}}J_0(\varrho)\label{eq:J0},
\end{equation}where $J_0$ is a Bessel function. The dominant contribution to the correlation function is an average over a circle of wavevectors with unit magnitude, and thus manifests as radially decaying ``oscillations''.

As a qualitative first account of finite-size effects, we may consider quantizing the wavevectors in the $x$-direction as $k_x=2\pi n/l_x$ while keeping $k_z$ continuous to simulate a highly anisotropic condensate; see Fig.~\ref{fig:gain}(a). We can still carry out steepest-descent integration of Eq.~(\ref{eq:g2d}) in the $k_z$ direction to obtain
\begin{equation}
        g_\perp(\x,\tau) \sim \frac{e^{2\tau}}{\sqrt{\tau}}\sum_{-1<k_x<1}\frac{\cos (\sqrt{1-k_x^2}z)}{\sqrt{1-k_x^2}}e^{ik_xx}\label{eq:cos}.
\end{equation}We recover Eq.~(\ref{eq:J0}) in the limit $l_x\rightarrow\infty$ by converting the sum over $k_x$ to an integral. In contrast, for $l_x\sim2\pi$ only a few $k_x$ are allowed, and the correlation function develops some anisotropy. As noted in the introduction, however, for any $k_x\leq1$ the gain in Eq.~(\ref{eq:omegak}) reaches the maximum $\Omega=1$ at some $k_z$ satisfying $k_x^2+k_z^2=1$, which implies that several wavevectors are equally dominant in their contribution to the correlation function. Moreover, it is clear from the denominator in Eq.~(\ref{eq:cos}) that the result is highly dependent on the allowed values of $k_x$, i.e., the exact value of $l_x$. As we show in the following sections, both of these predictions are incorrect. To adequately account for the effect of confinement, we now consider more rigorously the spin dynamics in the presence of a harmonic potential.

\subsection{Spectrum of $h_0$}
Working in dimensionful units for a moment, suppose the condensate is infinite in the $z$-direction but has a finite extent $L_x=2R$ in the $x$-direction. We model the confining potential as $V=\frac{1}{2}m\omega^2x^2$, with the frequency determined by $V(x=R)=\mu=c_0n_0$. Then in the Thomas-Fermi limit of vanishing kinetic contribution ($a_\text{osc}=\sqrt{\hbar/m\omega}\ll R$), the solution to Eq.~(\ref{eq:scalargp}) is given by
\begin{equation}
        \psi_0 = \lb n_0\lp 1 - \frac{x^2}{R^2} \rp\rb^{1/2}\label{eq:psi0}
\end{equation}
for $|x|<R$ and zero otherwise. Once $|x|>R$, the exact solution vanishes exponentially on the length
scale $\delta=(a_\text{osc}^4/2R)^{1/3}\ll R$ (Ref. \cite{pethick}). Correspondingly, the effective
potential in the Hamiltonian $H_0$ of Eq.~(\ref{eq:H0n}) grows rapidly for
$|x|>R$. This allows us to replace, in the limit $\delta\to 0$, the exact boundary conditions $\psi(x\rightarrow\pm\infty)=0$ for the eigenfunctions of $H_0$ with $\psi(x=\pm R)=0$ and simultaneously use Eq.~(\ref{eq:psi0}) for $\psi_0$ in Eq.~(\ref{eq:H0n}):
\begin{equation}
        H_0 = \frac{\hbar^2}{2m}\lb -\nabla^2 - \frac{R^2}{(R^2-x^2)^2}\rb.
\end{equation}
The divergence of the second term in the Hamiltonian at $x=\pm R$ is
regularized by the boundary condition $\varphi(x=\pm R)=0$.  We also
note that Eq.~(\ref{eq:scalargp}) can be solved directly by numerical
integration \cite{gp} to give a solution that is valid for nonzero
$\delta$, but we opt for this simpler solution in order not to obscure
the physics.

\begin{figure}
\includegraphics[width=3.2in]{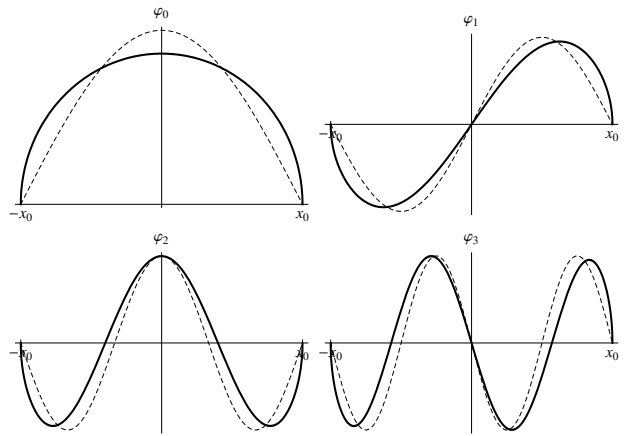}
\caption{\label{fig:phin} First four eigenstates of Eq.~(\ref{eq:phin}) (solid lines), along with eigenstates of the square-well problem (dashed).}
\end{figure}

Returning to dimensionless units, Eq.~(\ref{eq:H0}) becomes
\begin{equation}
        h_0 = -\nabla^2 - \frac{x_0^2}{(x_0^2-x^2)^2}\label{eq:h0}
\end{equation}where $x_0=R/\xi$ and the solutions obey $\varphi(x=\pm x_0)=0$. The value of $x_0$ controls the dynamics in the trapped case, which is highlighted by the fact that it may be written as a simple combination of the three energy scales in the system:
\begin{equation}
        x_0 = \frac{\sqrt{c_0n_0}\sqrt{|c_2|n_0}}{\frac{1}{2}\hbar\omega}.
\end{equation}The eigenstates and corresponding eigenenergies of this Hamiltonian are
\begin{equation}
        \varphi(\x) \propto e^{ik_zz}\varphi_n(x), \qquad \epsilon_{k_zn} = k_z^2 + \varepsilon_n
\end{equation}with $\varphi_n$ satisfying
\begin{equation}
        \lb -\ders{}{x} - \frac{x_0^2}{(x_0^2-x^2)^2}- \varepsilon_n\rb \varphi_n(x) = 0\label{eq:phin}
\end{equation}and $\int_{-x_0}^{x_0} dx\ |\varphi_n(x)|^2 = 1$. If we make the transformation
\begin{equation}
        \varphi_n(x) = \sqrt{1-\frac{x^2}{x_0^2}}\ \phi_n\lp\tanh^{-1} \frac{x}{x_0}\rp,
\end{equation}then the eigenvalue problem reduces to finding $K=x_0^2\varepsilon,\phi(\eta)$ such that
\begin{equation}
        \lb \ders{\phi}{\eta} + K\sech^4\eta \rb\phi(\eta) = 0
\end{equation}and $\phi(\eta\rightarrow\pm\infty)=\text{constant}$. This equation has been solved numerically; the first 4 eigenstates (converted back to $\varphi_n(x)$) and eigenvalues $K_n$ are shown in Fig.~\ref{fig:phin} and Table~\ref{tab:sech4}, respectively. Solutions to the square-well problem (width $2x_0$) obtained by neglecting the second term of Eq.~(\ref{eq:phin}) are also shown. The energy of the most relevant low-lying states differ significantly, illustrating the importance of accounting for the trap in a manner that properly preserves the zero mode.

We have already seen the resulting gain function in Fig.~\ref{fig:gain}(b), and it is worth noting that the presence of off-diagonal elements in $M$, Eq.~(\ref{eq:M}), leads to mode-mixing and anti-crossing behavior among curves of the same parity. Such crossings appear in the na\"{i}ve PBC quantization. There are remnants of the $\pm k_x$-degeneracy from the homogeneous spectrum, especially near $k_z\simeq0$. But most critical is the degree to which a single mode near $k_z\simeq1$ is emphasized in the resulting gain function, so that we expect a greater dominance of the corresponding wavelength in oscillations of the magnetization correlation function.

\begin{table}
\begin{tabular}{@{\hspace{.2cm}} c @{\hspace{.2cm}}| @{\hspace{.2cm}} c @{\hspace{.2cm}} | @{\hspace{.2cm}} c @{\hspace{.2cm}} | @{\hspace{.2cm}} c @{\hspace{.2cm}} | @{\hspace{.2cm}} c @{\hspace{.2cm}} }
$n$ & $0$ & $1$ & $2$ & $3$ \\ \hline
$K_n$ & $0.0$ & $4.62$ & $14.4$ & $29.1$  \\ \hline
$K_n^{\text{sw}}$ & 0.0& $7.40$ & $19.7$ & $37.0$  \\
\end{tabular}
\caption{\label{tab:sech4}First four eigenvalues for of Eq.~(\ref{eq:phin}) compared to the square-well (sw) problem obtained by ignoring the second term; $K_n=x_0^2\varepsilon_n$ is independent of the trap width. The bottom of the square well is taken to be at $-\pi^2/4$ so that the lowest energy is zero as required by $SU(2)$ symmetry.}
\end{table}

\subsection{Magnetization correlation function}

\begin{figure}
\includegraphics[width=3in]{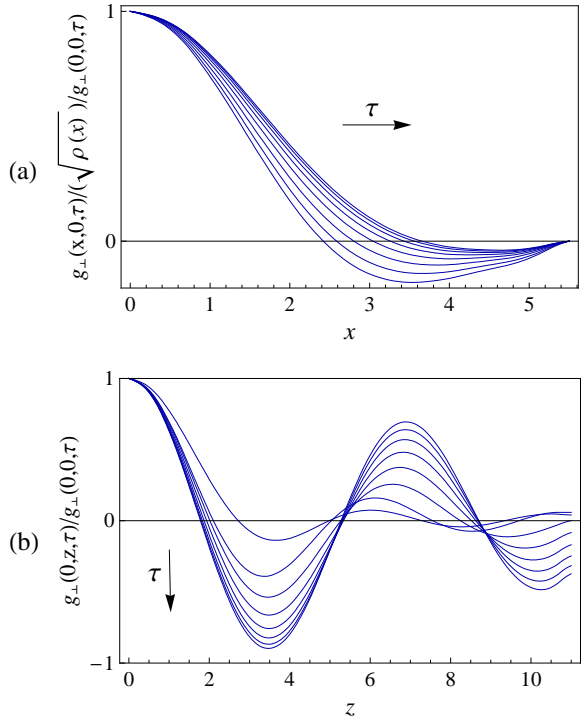}
\caption{\label{fig:slices} Plot of (a) $g_\perp(x,0,\tau)/\sqrt{\rho(x)}$ and (b) $g_\perp(0,z,\tau)$ divided by $g_\perp(0,0,\tau)$. Arrows indicate the progression of curves with increasing $\tau$. Note the different scales on the horizontal axes.}
\end{figure}

\begin{figure}
\includegraphics[width=3in]{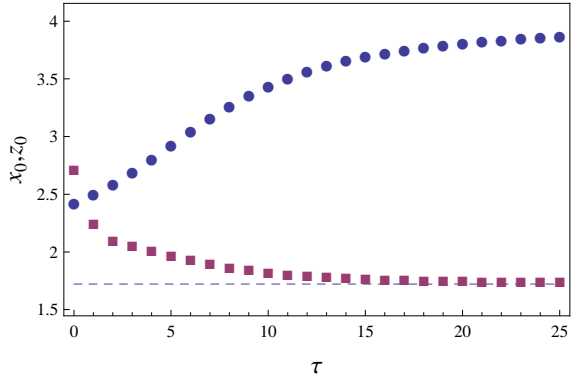}
\caption{\label{fig:nodes} (color online). The position of the first node, defined by $g_\perp(0,z_0,\tau)=0$ for the $z$-direction (squares) and $g_\perp(x_0,0,\tau)=0$ for the $x$-direction (circles), for the curves of Fig.~\ref{fig:slices}. The dashed line indicates the asymptotic value, 1.7, predicted for the $z$-direction from the point of maximum gain in Fig.~\ref{fig:gain}(b).}
\end{figure}

Based on the gain parameter for the unstable modes [Fig.~\ref{fig:gain}(b)], we claimed in the introduction the appearance of strong oscillations with a single wavelength in the $z$-direction. This is most dramatically illustrated by examining cross-sections of the correlation function at $x=0$ [Fig.~\ref{fig:slices}(b)], where it is clear that oscillations of the magnetization correlation function in the $z$-direction become stronger in time. The fact that this indicates the dominance of a \emph{single} mode (contrary to that predicted by PBCs) can be seen by plotting the positions of the first zero of $g_\perp(0,z,\tau)$ as shown in the lower curve of Fig.~\ref{fig:nodes}, which asymptotically approaches the value $\pi/2k_z^\ast,\ k_z^\ast=0.91$. The latter corresponds to the maximum of $\Omega(k_z)$ [belonging to curve 4 in Fig.~\ref{fig:gain}(b)]. We also show the first zero-crossing in the $x$-direction for comparison, although there is no meaningful wavelength in this direction and indeed the ``oscillations'' die out over time [Fig.~\ref{fig:slices}(a)]. This is easily explained in the homogeneous picture where the dominant mode occurs at $k_x^2+k_z^2=1$. Since $k_z\simeq1$, we expect the largest contribution in the $x$-direction to be $k_x\simeq0$, which has no oscillations.

\begin{figure}
\includegraphics[width=3in]{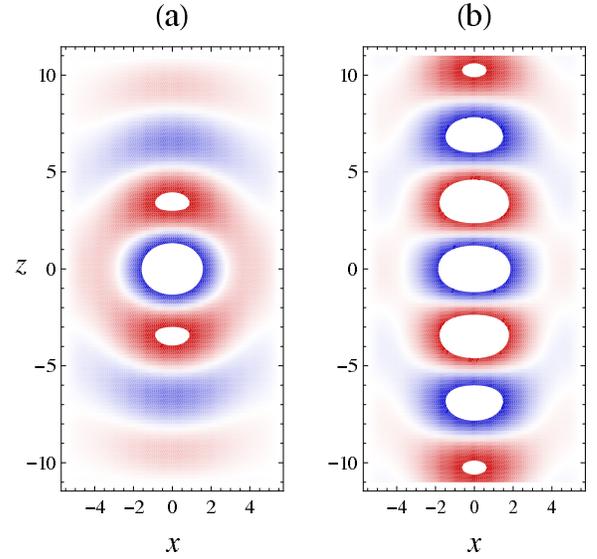}
\caption{\label{fig:trap} (color online). The density-normalized magnetization correlation function $g_\perp(0,\rho,\tau)/\sqrt{\rho(\x)}$ at (a) $\tau=4$ and (b) $\tau=15$. Dark regions represent alternating negative (red online) and positive (blue online) correlations, with strongest positive correlation at the center; the color scale is arbitrary. Large correlations were truncated to white to maximize contrast in the rest of the plot.}
\end{figure}

As we show in the next section, the underlying equations of motion are valid only until $\tau\simeq4$. Meanwhile the asymptotic behavior is not realized until $\tau\simeq 15$, as can be seen in Fig.~\ref{fig:nodes}. This can also be predicted on the basis of the gain curves in Fig.~\ref{fig:gain}(b) by noting that the maximum gain of the top curve is separated from the maximum gain of the curve below by approximately 0.2; thus $\tau=15$ represents roughly 3 time constants. Still, we expect even at fairly short times for the anisotropy to be visible in the full correlation function as shown in Fig.~\ref{fig:trap}, where we plot the correlation function $g_\perp(0,\x,\tau)/\sqrt{\rho(\x)}$ in the $x-z$ plane. Note that, since we are interested in the spin correlation, we have factored out $\sqrt{\rho}$. Smaller values of $x_0$ accelerate the development of anisotropy so that it may be conclusively discerned in a real experiment before depletion and nonlinear effects begin to play a significant role.

\subsection{Depletion of the polar phase}

\begin{figure}
\includegraphics[width=3in]{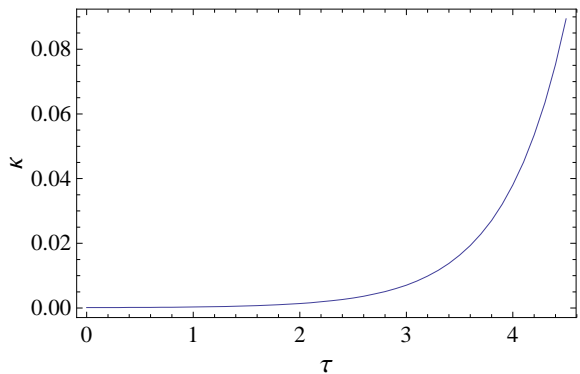}
\caption{\label{fig:depletion} The depletion function $\kappa$. Each unit of time represents approximately 16 ms. We have chosen $l_x=11,\ l_z=140$, close to experimental conditions \cite{stamper-kurn:2}.}
\end{figure}

In this section we consider for how long the underlying Eqs.~(\ref{eq:p1}),(\ref{eq:m1}), which were based on a zero-depletion approximation, are valid. Neglecting thermal effects, the number of $m_F=0$ particles taken out of the condensate, or equivalently the number of $m_F=\pm1$ pairs created, is given by $N_\text{pairs}(\tau) = \int d\x\ n_\text{pairs}(\x,\tau)$ where
\begin{align}
        n_\text{pairs}(\x,\tau) &= \mean{\hat{\Phi}_{+1}^\dag(\x,\tau)\hat{\Phi}_{+1}(\x,\tau)
        + \hat{\Phi}_{-1}^\dag(\x,\tau)\hat{\Phi}_{-1}(\x,\tau)}  \\
        &= 2\sum_{\alpha,\beta,\gamma}\varphi_\alpha(\x)\varphi^\ast_\beta(\x)v_{\alpha\gamma}(\tau)v^\ast_{\beta\gamma}(\tau).
\end{align}Rather than evaluate what turns out to be a cumbersome expression, we note from Eqs.~(\ref{eq:lincomb}) and (\ref{eq:gperp}) that $n_\text{pairs}(\x,\tau) \leq g_\perp(\x,\x,\tau)/\rho(\x)$. The expression is well-approximated by evaluating the integrand at $\x=0$, so that the condensate fraction $\kappa=N_\text{pairs}/N$ is given by
\begin{equation}
        \kappa(\tau) = \frac{g_\perp(0,0,\tau)}{\xi^dn_0^{(d)}}\label{eq:kappa},
\end{equation}where $d$ is the dimensionality of the system. In the case of a quasi-2D trap, $d=2$ and $n^{(2)}_0=\int dy\ n(0,y,0)$. The condition for the governing equations, and hence the resulting magnetization correlation function, to be a good description of the system is therefore $\kappa\ll1$. Since the denominator in Eq.~(\ref{eq:kappa}) represents the number of particles in a spin correlation volume, physically this condition requires the magnetization per particle in each individual domain to be small. Fig.~\ref{fig:depletion} plots the depletion as a function of time, showing that the zero-depletion approximation is quite good until $\tau^\ast\simeq 4$ (i.e., $t^\ast\simeq64\unit{ms}$) for the conditions of \cite{stamper-kurn:2}, after which the nonlinear effects not taken into account in the equations of motion are expected to play a significant role in the dynamics. The difference in gain $\Delta\Omega\simeq 1/\tau^\ast$ that can be resolved in the exponential growth of the different modes before depletion renders the underlying equations invalid is indicated by the length of the arrow in Fig.~\ref{fig:gain}(b).

\section{Conclusions}
We have analyzed the spin dynamics of a spatially anisotropic spinor BEC starting from the polar ($m_F=0$) state, which is unstable in the absence of a high magnetic field and develops ferromagnetic domains. We find that anisotropy should develop in the magnetization correlation function in the form of strong oscillations in the unconfined direction. We have ignored the effects of dipole-dipole interactions, which are inherently anisotropic but are an order-of-magnitude weaker than the spin-spin interactions \cite{saito:dipolar}. Although dipolar effects can be important under certain circumstances \cite{stamper-kurn:3,demler}, preliminary investigation of dipolar forces in the system described here suggests that they will not significantly alter the magnetization correlation anisotropy. Nevertheless, our findings are contrary to what has been observed experimentally \cite{stamper-kurn:2}, which poses the question of what is causing the observed preference for the trapped direction. An investigation of the possible effects of optical aberration in the imaging device used in the experiment is underway, and the new experiments may shed light on the origin of the presently evident inconsistency \cite{stamper-kurn:pc}.

\section*{ACKNOWLEDGMENTS}
We are grateful to D. M. Stamper-Kurn for helpful discussions. This work was supported by NSF Grant No.~DMR-0603369 and DOE Grant No.~DE-FG02-08ER46482.

\bibliographystyle{h-physrev}

\end{document}